\documentclass[10pt,journal]{IEEEtran}

\usepackage[dvips]{graphicx}

\usepackage{url}

\usepackage{fix-cm}

\usepackage[T1]{fontenc}
\usepackage[utf8]{inputenc}
\usepackage{amsmath} 
\usepackage{amsfonts} 
\usepackage{amssymb} 
\usepackage{amsthm}
\usepackage{color}
\usepackage{url}
\usepackage{subcaption}
\usepackage{float}
\usepackage[linesnumbered,ruled,vlined]{algorithm2e}
\usepackage{cleveref}
\usepackage{verbatim}
\usepackage{multirow}
\usepackage{comment}
\SetKwInput{KwInput}{Input}                
\SetKwInput{KwOutput}{Output}              
\SetKwInput{Kwinitialize}{Initialization}              
\usepackage{tikz}
\usepackage{nicematrix}

\renewcommand{\vec}[1]{{\bf{#1}}} 
\newcommand{\vecgreek}[1]{{\boldsymbol{#1}}} 
\newcommand{\tran}{^{\mbox{\scriptsize T}}}
\newcommand{\herm}{^{\mbox{\scriptsize H}}}

\DeclareMathOperator{\Tr}{Tr}
\newcommand{\diag}{\mathrm{diag}}

\DeclareMathOperator*{\argmin}{argmin}
\DeclareMathOperator*{\argmax}{argmax}

\newcommand{\gammab}{\vecgreek{\gamma}}

\newcommand{\Sigmab}{\vecgreek{\Sigma}}
\newcommand{\lambdab}{\vecgreek{\lambda}}

\newcommand{\mybibliography}{\bibliography{conf_short,jour_short,Bi.bib}}

\title{Adaptive and Self-Tuning SBL with Total Variation Priors for Block-Sparse Signal Recovery}

\author{
\IEEEauthorblockN{Hamza Djelouat, Reijo Leinonen,  Mikko J. Sillanpää,  Bhaskar D. Rao, ~\IEEEmembership{Life Fellow,~IEEE} and Markku Juntti,~\IEEEmembership{Fellow,~IEEE}}
\thanks{Hamza Djelouat and Markku Juntti  are with Centre for Wireless Communications -- Radio Technologies, FI-90014, University of Oulu, Finland}
\thanks{ Reijo Leinonen and  Mikko J. Sillanpää,  are with Research Unit of Mathematical Sciences, University of Oulu, Finland., University of Oulu, Finland}
\thanks{Bhaskar D. Rao is with the Department of Electrical and Computer Engineering, University of California, San Diego, La Jolla, CA 92093, USA }
\thanks{This work has been supported  by the Research Council of Finland (6GWiCE project, grant 319485) and (6G Flagship program, grant 369116). Bhaskar D. Rao’s work was supported by NSF (grant 
2225617).}
}

\begin{document}
\maketitle
\begin{abstract}
This letter addresses the problem of estimating block sparse signal with unknown group partitions in a multiple measurement vector (MMV) setup. We propose a Bayesian framework by applying an adaptive total variation (TV) penalty on the hyper-parameter space of the sparse signal. The main contributions are two-fold. 1) We extend the TV  penalty beyond the immediate neighbor, thus enabling better capture of  the signal structure.  2) A dynamic framework is provided to learn the penalty parameter for regularization. It is based on the statistical dependencies between the entries of tentative blocks, thus eliminating the need for fine-tuning. The superior performance of the proposed method is empirically demonstrated by extensive computer simulations with the state-of-art benchmarks. The proposed solution exhibits both excellent performance and robustness against sparsity model mismatch.

\end{abstract}
%


%
\vspace{-.2cm}
\section{Introduction}

Block sparsity, where nonzero elements occur in clusters rather than uniformly in the data, provides a powerful framework for signal recovery. Such a grouping  pattern arises in many practical scenarios, such as near-field and millimeter-wave channel estimation \cite{Cui2022channel, Li2018millimeter}, and user activity detection in machine-type communications \cite{djelouat2024hierarcichal}. While block sparsity provides valuable prior knowledge about the signal structure, leveraging this information presents a key challenge. The recovery process must simultaneously determine the block locations, sizes, and values, making it substantially more complex than conventional sparse signal recovery where only individual nonzero elements need to be identified.



Sparse recovery algorithms developed under  Compressive sensing (CS) \cite{Candes-Romberg-Tao-06,Donoho-06}  framework excel at promoting sparsity but ignore the inherent block structures present in many signals. For scenarios where block boundaries are known \textit{a priori}, the CS algorithms have been adapted to exploit this structure. Those include block compressive sampling matching pursuit (block-CoSaMP) \cite{baraniuk2010model}, block orthogonal matching pursuit (block-OMP) \cite{Eldar2010block}, and block-sparse Bayesian learning (BSBL) \cite{Zhang2013Extension}. By incorporating signal-specific block structures, these recovery algorithms achieve significantly better performance than their conventional counterparts. 

The  practical scenarios present an additional challenge of the \emph{unknown block boundaries and sizes}. This has spurred development of adaptive techniques that simultaneously promote block sparsity and detect boundaries, primarily through modeling inter-element dependencies within the signal.

Sparse signal recovery with unknown boundaries has been approached through two main frameworks: Bayesian and regularization-based methods. For instance, pattern-coupled (PC) sparse Bayesian learning (SBL)  and its variants \cite{fang2014pattern,Dai2019non,8319524,WANG2020107255} promote block sparsity by hard-coupling neighboring elements in the signal's variance vector. Meanwhile, regularization approaches such as fused least absolute shrinkage and selection operator (LASSO) \cite{nowak2011fused} employs total variation (TV) penalty to discourage rapid signal fluctuations. A recent work \cite{Aditya2022block} combined these two approaches by applying the TV regularization to the signal hyper-parameters within an SBL framework, effectively reducing transitions between zero and non-zero regions. However, all of these frameworks have two key limitations: they restrict smoothing to the immediate neighbor only, overlooking broader structural patterns, and require careful tuning of the block-sparsity promoting term -- a challenge when block sparsity varies \cite{Aditya2022block}.


This letter introduces a novel solution to block-sparse signal recovery by formulating the problem using an SBL cost function augmented by a \emph{log-based TV} penalty with  \emph{learnable} weights. Specifically, the proposed approach aims to (i) extend the hyper-parameter coupling from a single neighbor to multiple consecutive neighbors, thereby enhancing the model's ability to capture local structures; and (ii) to provide a framework to dynamically and automatically learn the regularization parameter based on the signal statistics on each iteration. Finally, to ensure computational efficiency, we employ an alternating direction method of multipliers (ADMM) framework, which allows for closed-form updates at each step of the optimization process.  We validate the effectiveness of the proposed approach via extensive experiments, showing significant improvements in signal recovery performance over existing state-of-the-art methods in terms of quality and robustness to model mismatch.

\section{Block-Sparse Recovery Problem Overview}
\subsection{System Model}
We consider the estimation of the unknown block-sparse signal ensemble ${\vec{X}=[\vec{x}_1,\ldots,\vec{x}_N]\in \mathbb{C}^{N \times M}}$ 
from a set of  noisy linear measurements $\vec{Y}=[\vec{y}_1,\ldots,\vec{y}_M]\in \mathbb{C}^{L\times M}$ as
\begin{equation}\label{y}
\vec{y}_m=\vec{A}\vec{x}_m+\vec{w}_m, \quad m=1,\ldots,M,    
\end{equation}
where ${\vec{A} \in \mathbb{C}^{L\times N}}$ is a fixed and known measurement matrix, and each ${\vec{w}_m \sim \mathcal{CN}(\vec{0},\sigma^2\vec{I}_L) }$ is the additive noise vector.

We focus on a signal ensemble that exhibits two key properties: (i) a \emph{row-sparse} structure, meaning all signals $\vec{x}_m$, $\forall m$, share the same support, and (ii) a \emph{block-sparse} structure, where each signal $\vec{x}_m$, $\forall m$, comprises an \emph{unknown} number of blocks with \emph{varying} sizes. 
The unknown signal $\vec{x}_m$ is modeled to follow a multivariate complex Gaussian distribution as
\begin{equation}\label{eq:x_gauss}
    \vec{x}_m \sim \mathcal{CN}(\vec{0},\diag(\gammab)), \quad m=1,\ldots,M
\end{equation}
where ${{\gammab=[\gamma_1,\ldots,\gamma_N]\tran \in \mathbb{R}_+^{N}}}$ represents the \emph{unknown hyper-parameter vector}  controlling the sparsity of each $\vec{x}_m$.

\subsection{SBL Framework}
Given the system model in \eqref{y}, the likelihood function is Gaussian or
${p(\vec{y}_m|\vec{x}_m;\sigma_{\mathrm{w}})= 
\mathcal{CN} (\vec{A} \vec{x}_m, \sigma^2_{\mathrm{w}})}$. Thus by using the conjugate Gaussian prior in \eqref{eq:x_gauss} along the likelihood of Gaussian density, the  posterior density $p(\vec{x}_m|\vec{y}_m; \gammab)$ is also Gaussian, given as
\begin{equation}
    p(\vec{x}_m|\vec{y}_m; \gammab)\sim \mathcal{CN}(\mu_{\vec{x}_{m}},\Sigmab_{\vec{x}}),
\end{equation}
where 
\begin{equation}
\mu_{\vec{x}_m}=\frac{1}{\sigma^2}\Sigmab_{\vec{x}}\vec{A}\herm\vec{y}_m, \quad  \Sigmab_{\vec{x}}=(\frac{1}{\sigma^2}\vec{A}\herm\vec{A}+\diag(\gammab)^{-1})^{-1},
\end{equation}
For known $\gammab$, the estimated signal is given as $\vec{X}=[\mu_{\vec{x}_1},\ldots,\mu_{\vec{x}_M}]$.
The SBL framework aims to estimate the hyper-parameter vector $\gammab$  via Type-II maximum \textit{a posteriori} (MAP) estimation as
\begin{equation}\label{sbl}
\begin{array}{ll}
      \hat{\gammab}&= \displaystyle  \argmax_{\gammab\succeq 0} \log p(\gammab|\vec{Y})\\
   &  = \displaystyle  \argmin_{\gammab\succeq 0} \log \det\big( \Sigmab_{\mathrm{y}}\big)+ \Tr[\vec{Y}\Sigmab_{\vec{y}}^{-1}]-\log p(\gammab),
\end{array}
\end{equation}
where $\Sigmab_{\mathrm{y}}=\sigma_{\mathrm{w}}^2 \vec{I}+\vec{A}\diag(\gammab)\vec{A}\herm$ is the measurement model covariance matrix and $\log p(\gammab)$ is the \emph{hyper-prior} on $\gammab$.  The problem \eqref{sbl} is non-convex due to the logdet term. Therefore, several approaches have been proposed to approximate the solution such  via block coordinate descent \cite{Fengler2021non}, fixed point iteration \cite{tipping2001sparse}, and expectation maximization (EM) \cite{wipf2004sparse,zhang2011sparse}. We will employ the latter to solve the problem.

The EM-SBL iterates between two steps as follows

\textbf{E-step}: The joint distribution is given as
 \begin{equation}
p(\vec{Y},\vec{X},\gammab)=p(\vec{Y}|\vec{X},\gammab)p(\vec{X}|\gammab)p(\gammab).
 \end{equation}
At any particular iteration $(k)$, the E-step with respect to $\gammab$ is evaluated by averaging out the hidden variable $\vec{X}$ as
\begin{equation}\label{eq:Q_theta}
    \begin{array}{ll}
Q(\gammab^{k}|\gammab^{k-1})&=\mathbb{E}_{p(\vec{X}|\vec{Y},\gammab^{(k-1)})} \bigg[\log p(\vec{Y}|\vec{X},\gammab)p(\vec{X}|\gammab)p(\gammab) \bigg]\\&=\mathbb{E}_{p(\vec{X}|\vec{Y},\gammab^{(k-1)})} \bigg[\log p(\vec{X}|\gammab)+\log p(\gammab)\bigg]\\&\displaystyle\propto -\sum_{i=1}^N \Bigg[M\log (\gamma_i)+\frac{s_i^{(k-1)}}{\gamma_i}\Bigg]+\log p(\gammab),\\
    \end{array}
\end{equation}
where $s_{i}^{(k-1)}=\Sigmab_{i,i}^{(k-1)}+\frac{1}{M}\|\mu_x^{(k-1)}\|^2$.

\textbf{M-step}: The hyper-parameter is updated as follows
\begin{equation}\label{M-step}
       \hat{\gammab}=\displaystyle \argmin_{\gammab}- Q(\gammab^{k}|\gammab^{k-1})=\displaystyle \argmin_{\gammab}f(\gammab)-\frac{1}{M}\log p(\gammab),
\end{equation}
where 
\begin{equation}
f(\gammab)=\sum_{i=1}^N f(\gamma_i)=\sum_{i=1}^N  \log(\gamma_i)+\frac{s_i}{\gamma_i}.
\end{equation}
To promote block-sparse solutions in the EM-SBL framework, it is essential to employ \textbf{prior functions} that simultaneously encourage sparsity and capture the inherent block structures in the signal. In the existing literature, block sparsity is typically enforced through two main approaches: 1) a hard prior coupling scheme where each hyper-parameter is defined as $\gamma_i\triangleq f(\gamma_i,\gamma_{i+1},\gamma_{i-1},\beta)$, as seen in PC-SBL inspired solutions, and 2) a soft prior formulation given by $-\log p(\gammab)=\beta\sum_{i=2}^N |\log(\gamma_i)-\log(\gamma_{i-1})|$, where $\beta>0$ is a regularization parameter controlling the coupling strength in both approaches. However, these approaches suffer from two major limitations: they only consider immediate neighboring elements in their coupling mechanisms, and they require careful tuning of the parameter $\beta$, making them susceptible to sparsity model mismatches, particularly when isolated non-zero elements appear in the signal.



We aim to improve the existing solutions and to bridge some gaps to overcome the limitations of the proposed approaches in terms of accuracy and robustness of the estimation. 
We will first discuss the design of the prior function, followed by the detailed derivation of the solution for the M-step.

\section{ Proposed Solution }
Before proceeding to solve problem \eqref{M-step}, we first discuss the appropriate choice of $p(\gammab)$ to promote a \emph{robust structured} sparse solution.  Our approach builds upon the framework introduced in  \cite{Aditya2022block} and enhances the SBL inference framework  through the following key ideas. 1)
\textbf{Neighborhood smoothing} extends the hyper-parameter coupling from a single neighbor to multiple consecutive neighbors, hence,  improving the model's ability to adapt to the signal structure and capture local dependencies between the elements of the signal. 2)\textbf{ Automatic regularization tuning} by learning the set of  regularization parameters$\{\beta_{i,j}\}$, where each $\beta_{i,j}$ is unique for each pair of $\gamma_i$ and $\gamma_j$. This is achieved by adaptively updating $\{\beta_{i,j}\}$ at each EM iteration based on the signal statistics as we will show later. 
\vspace{-.3cm}
\subsection{Design of Hyper-prior $p(\gammab)$}
To this end, we propose the following hyper-prior function: 
\begin{equation}\label{log(p)}
-\log p(\gammab) = \sum_{i=2}^N\sum_{j=1}^{i-1} \beta_{i,j} |g(\gamma_i) - g(\gamma_j)|, \end{equation}
where $\beta_{i,j} \in [0,1]$ is a weight parameter, and $g(\cdot)$ is a function that measures the number of transitions between zero and non-zero regions. Ideally, we would choose $g(\cdot) = \mathbb{I}(\cdot)$ (the indicator function), but this would result in a binary optimization problem that is computationally intractable. Instead, we use the approximation $g(\cdot) = \log(\cdot)$, as the log function provides a suitable approximation to the counting function $\mathbb{I}(\cdot)$ \cite{candes2008enhancing}.

This letter proposes to use a set of $\{\beta_{i,j}\}$ for each pair of $\{\gamma_i, \gamma_j\}$ as
\begin{equation}
    \beta_{i,j}=\frac{1}{Z_i} \psi\bigg[d\big(\log(\gamma_i),\log(\gamma_j)\big)\bigg],
\end{equation}
where $Z_i$ is a normalization factor that ensures
that the weights sum to 1, function $\psi[\cdot]$ is a kernel decay function and
$d(\cdot)$ is a distance function that measures the similarity between the
 sparsity profiles of the $i$th and $j$th  elements. Both functions $\psi(\cdot)$ and $d(\cdot)$ should be designed in a way to set a zero penalty on the smoothing between $i$th and $j$th elements if their sparsity profile is different,  while setting a positive weight when the $i$th and the $j$th elements  have a similar sparsity profile. To this end, we set $\psi$ as a negative exponential and set $d(\cdot)$  as an Euclidean norm function. Subsequently,  we define  $\beta_{i,j}$ at each EM iteration $(k)$ as follows
\begin{equation}
    \beta_{i,j}^{(k)}=
    \begin{cases}
        \exp{( -\|\log(\gamma_i^{(k-1)})-\log(\gamma_j^{(k-1)})\|^2)},& \text{if $j \in \Omega_i$} \\
      0 & \text{otherwise},
    \end{cases}
\end{equation}
where $\Omega_i$ is a set of the closest neighbors of the $i$th element to be selected empirically.

\subsection{ADMM Solution for the M-step}
We use the proposed $p(\gammab)$ and express the optimization problem for the M-step at the $k$th EM iteration as\footnote{We omit the EM iteration index in this subsection for sake of simplicity.}
\begin{equation}\label{M-step_general}
    \begin{array}{cc}
         \hat{\gammab}=& \displaystyle \arg \min_{\gammab} f(\gammab)+\frac{1}{M}\sum_{i=1}^N\sum_{j=1}^{i-1} \beta_{i,j}|\log(\gamma_{i})-\log(\gamma_{j})|.  
    \end{array}
\end{equation}
We propose an ADMM framework  \cite{boyd2011distributed}  to solve the M-step \eqref{M-step_general}.  
First, let us introduce the auxiliary variable $\vec{C} \in \mathbb{R}^{N\times N}$ and we write the M-step as follows
\begin{equation}\label{Mstep}
    \begin{array}{ll}
    & \displaystyle \argmin_{\gammab,\vec{C}} f(\gammab)+\frac{1}{M}\sum_{i=2}^N\sum_{j=1}^{i-1} |C_{i,j}|\\  &\mbox{s.t.} \,\, C_{i,j}=\beta_{i,j}\big(\log(\gamma_{i})-\log(\gamma_{j})\big),\forall i,j \in \mathcal{N}.
    \end{array}
\end{equation}
The  problem in \eqref{Mstep} refactors the original M-step \eqref{M-step_general} into a bi-convex optimization problem over $\gammab$ and $\vec{c}$. Thus,  providing an excellent framework to apply an alternating optimization technique such as ADMM.  Subsequently, we write the augmented Lagrangian function as
\begin{equation}\label{Lagrange_Mstep}
         \mathcal{L}(\gammab,\vec{c})=\displaystyle  f(\gammab)+\sum_{i=2}^N\sum_{j=1}^{i-1}\bigg[  \frac{1}{M}|C_{i,j}|\displaystyle+\frac{\rho}{2} \|C_{i,j}-\bar{\gamma}_{i,j}+\frac{\lambda_{i,j}}{\rho}\|^2\bigg], 
\end{equation}
where  ${\bar{\gamma}_{i,j}=\beta_{i,j}\Big(\log(\gamma_i)-\log(\gamma_j) \Big)}$, $\lambdab \in \mathbb{R}^{N\times N}$ denotes the dual variable matrix, and $\rho$ is a real small positive variable. 

The ADMM framework solves \eqref{M-step_general} by iteratively minimizing $\mathcal{L}(\cdot)$ over the primal variables $(\gammab,\vec{C})$, followed by updates to the dual variables $\lambdab$ \cite{boyd2011distributed}. We detail the derivation of each ADMM iteration step $(t)$ below.

\subsubsection{$\gammab$-update}
First, we update each element of the  hyper-parameter vector sequentially as 
\begin{equation}\label{gammab_k+1_obj}
\begin{array}{ll}
     \gamma_i^{(t+1)}&=\displaystyle \argmin_{\gamma_i} \log (\gamma_i)+ \frac{\rho}{2}\sum_{\substack{j \in \Omega_i\\ j<i}}\|\tilde{a}_{j}^{(t)}-\beta_{i,j}\log(\gamma_{i})\|^2\\& \displaystyle+ \frac{\rho}{2}  \sum_{\substack{j \in \Omega_i\\ j>i}}|\bar{a}_{j}^{(t)}+\beta_{i,j}\log(\gamma_{i})\|^2+\frac{s_i}{\gamma_i},
\end{array}
\end{equation}
where $\tilde{a}_{j}^{(t)}= C_{i,j}^{(t)}+\beta_{i,j}\log(\gamma_{j}^{(t)})+\frac{\lambda_{i,j}^{(t)}}{\rho}$ and $\bar{a}_{j}^{(t)}= c_{j,i}^{(t)}-\beta_{i,j}\log(\gamma_{j}^{(t)})+\frac{\lambda_{j,i}^{(t)}}{\rho}$. 
The solution can be obtained by setting the gradient of the objective function in \eqref{gammab_k+1_obj} to zero
\begin{equation}\label{gammab_k+1}
    \gamma_i^{(t+1)}=\frac{s_i}{1+\displaystyle \rho\sum_{j \in \Omega_i} \beta_{i,j}-\rho\bigg(\sum_{ j\in \Omega_i,j<i} \tilde{a}_{j\\j}^{(t)}-\sum_{j\in \Omega_i,j>i} \bar{a}_{j}^{(t)}\bigg)}.
\end{equation}
\subsection{$\vec{C}$-update}
The optimization problem with respect to $\vec{C}$ can be decomposed into $N(N-1)/2$ convex subproblems as follows
\begin{equation}\label{c_k+1_obj}
    C_{i,j}^{(t+1)}=\displaystyle \argmin_{C_{i,j}}  \frac{1}{M}|C_{i,j}|+\frac{\rho}{2}\|C_{i,j}-\bar{\gamma}_{i,j}^{(t+1)}+\frac{\lambda_{i,j}^{(t)}}{\rho}\|, 
\end{equation}
where  $\bar{\gamma}_{i,j}^{(t+1)}=\beta_{i,j}\big(\log(\gamma_i^{(t+1)})-\log(\gamma_j^{(t+1)}) \big)$. The expression in \eqref{c_k+1_obj} admits a closed-form solution via a proximal
gradient method \cite{goldstein2014field} as
\begin{equation}\label{c_k+1}
    C_{i,j}^{(t+1)}=\eta_{\frac{1}{M\rho}}(\bar{\gamma}_{i,j}^{(t+1)}-\frac{\lambda_{i,j}^{(t)}}{\rho}),
\end{equation}
where $\eta_\kappa(a)=\text{sign}(a)\max(|a|-\kappa)$ is a soft-thresholding operator for any $ a, \kappa\in \mathbb{R}_+$.
\subsection{$\lambda$-update}
Finally, the dual variable is update as
\begin{equation}\label{lambda_k+1}\footnotesize
\lambda_{i,j}^{(t+1)}=\lambda_{i,j}^{(t)}+\rho\bigg(C_{i,j}^{(t+1)}-\bar{\gamma}_{i,j}^{(t+1)}\bigg), \forall i,j
\end{equation}

The proposed solution is summarized as Algorithm 1.
\begin{algorithm}[t]
\DontPrintSemicolon
   \footnotesize \KwInput{$\vec{A}$, $\vec{Y}$,
    $\epsilon$, $\rho$, $t_{\mathrm{max}}$, $k_{\mathrm{max}}$, $\{\Omega_i\}_{i=1}^N$.}
\footnotesize \Kwinitialize{$ \gammab^{(0)},\vec{C}^{(0)},\lambdab^{(0)}, {t=1, k=1.}$}

   \While{$k<k_{max}$ $\mathrm{or}$ $\| \vecgreek{\mu}_\mathrm{x}^{(k)}-\vecgreek{\mu}_\mathrm{x}^{(k+1)} \|^2>\epsilon$ }
   {
$\Sigmab_{\vec{x}}^{(k+1)}=\Bigg(\frac{1}{\sigma^2}\vec{A}\herm\vec{A}+\diag(\gammab^{(k)^{-1}})\bigg)^{-1}$\\
 
$\mu_{\vec{x}}^{(k+1)}=\frac{1}{\sigma^2}\Sigmab_{\vec{x}}^{(k+1)}\vec{A}\herm\vec{y}$\\
  $\beta_{i,j}=\exp\big(-\|\log(\gamma_i^{(k)})-\log(\gamma_j^{(k)})\|^2))$, $\forall i, \forall j $\\

  \While{\scriptsize$t<t_{\mathrm{max}}$ or $\|\gammab^{(t)}-\gammab^{(t+1)}\|^2>\epsilon$}
   {
    
 Update $\gammab^{(t+1)}$ using \eqref{gammab_k+1}\\
 Update $\vec{C}^{(t+1)}$ using \eqref{c_k+1}\\
Update $\lambdab^{(t+1)}$ using \eqref{lambda_k+1}\\
$\scriptsize t \leftarrow{t+1}$\;
 }
$\scriptsize k \leftarrow{k+1}$\;
 
}
\caption{The proposed algorithm}
\end{algorithm}

\section{Simulation Results}
\begin{figure*}[h]\centering
\subfloat[]{\includegraphics[scale=0.35]{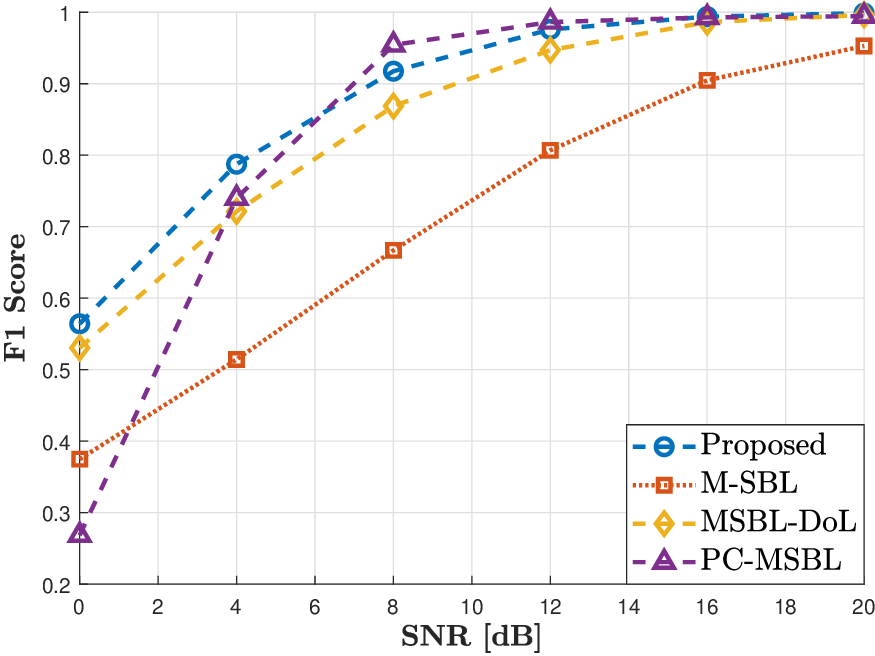}}\hfil\unskip 
\subfloat[]{\includegraphics[scale=0.35]{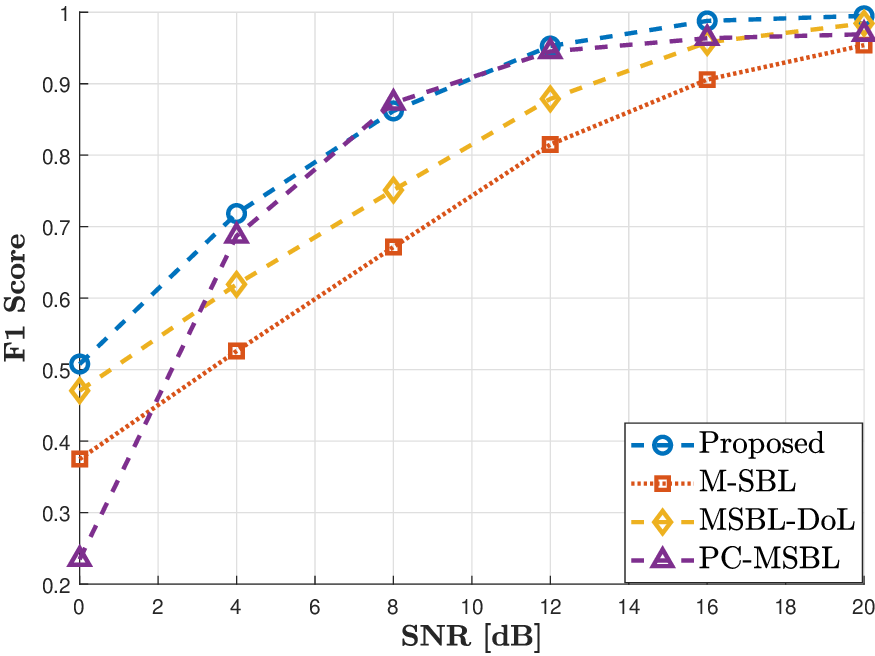}}\hfil\unskip
\subfloat[]{\includegraphics[scale=0.35]{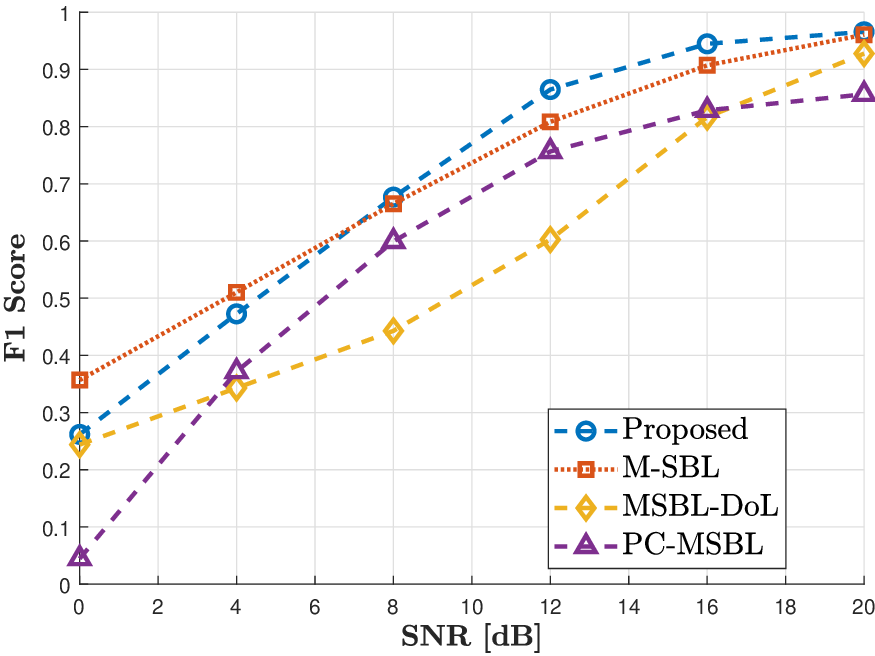}}
\caption{Comparison of F1-score under varying signal-to-noise-ratio (SNR) levels and sparsity patterns: (a) block sparsity, (b) hybrid sparsity, and (c) isolated non-zero elements.}\label{F1}

  \end{figure*}

\begin{figure*}\centering
\subfloat[]{\includegraphics[scale=0.35]{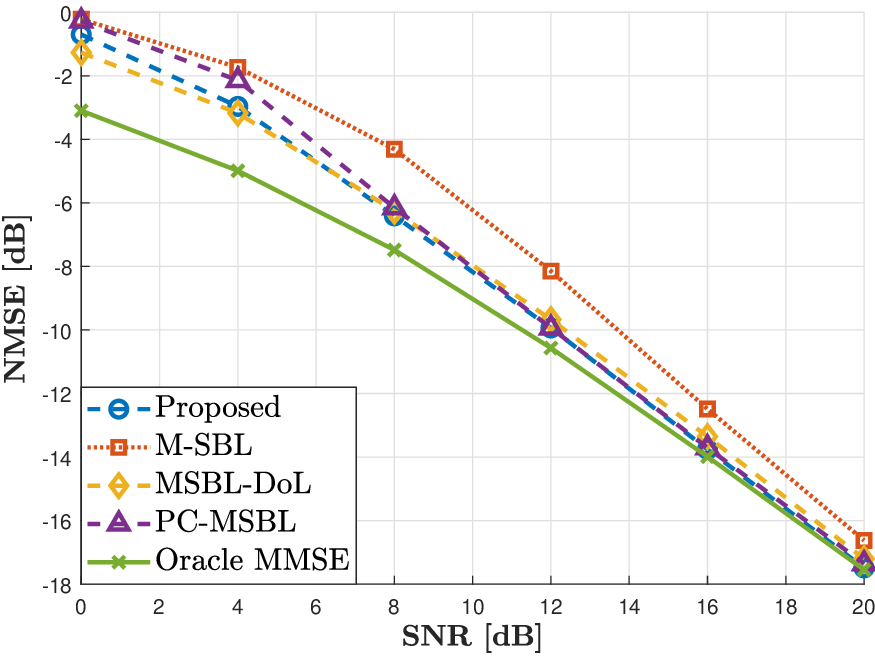}}\hfil\unskip 
\subfloat[]{\includegraphics[scale=0.35]{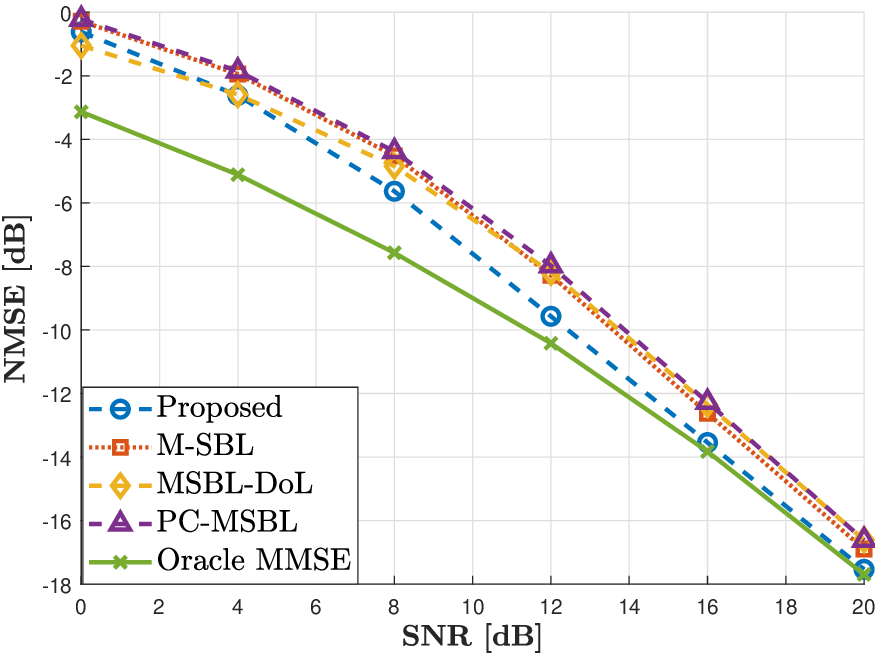}}\hfil\unskip
\subfloat[]{\includegraphics[scale=0.352]{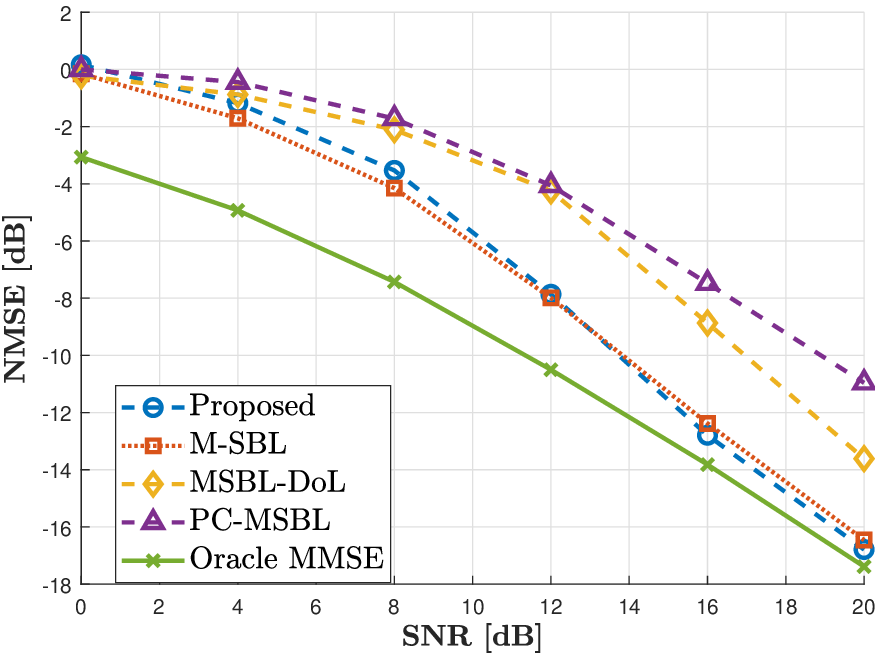}}
\caption{Comparison of the signal reconstruction quality in terms of NMSE under varying SNR levels and sparsity patterns: (a) block sparsity, (b) hybrid sparsity, and (c) isolated non-zero elements.}\label{MSE}
  \end{figure*}

We consider a system setup with $N = 300$, $M=5$,  $L=30$ and   $\vec{A}$ drawn from an i.i.d.\ complex Bernoulli distribution with normalized columns. The performance is evaluated in terms of the normalized mean
square error (NMSE) given as $\frac{\mathbb{E}\left [\| \vec{X}-\hat{\vec{X}}\|_{\mathrm{F}}^2 \right ]}{\mathbb{E}\left[\| \vec{X}\|_{\mathrm{F}}^2 \right]}$, and 
the  support recovery  rate  using the F1-Score $F1=\frac{\text{precision}\times \text{recall}}{\text{precision}+\text{recall}}$ as defined in \cite{Aditya2022block}.

We compare the proposed algorithm with three baseline sparse recovery algorithms: 1) M-SBL \cite{wipf2007empirical}, 2) MSBL-DoL \cite{Aditya2022block}, 3) PC-SBL \cite{fang2014pattern}. In addition, we include the oracle minimum mean square error (MMSE) estimator to provide a baseline for NMSE performance.

In order to evaluate the performance of the proposed solution under  varying signal sparsity structures, we examine three use cases: 1) \emph{block sparsity} : 5 non-zero blocks, each containing 5 elements, 2) \emph{hybrid sparsity}: 3 active blocks, each containing 5 elements in addition to 5 randomly selected elements, and 3) \emph{random activity}: 25 non-zero elements  randomly selected non-zero elements.


Figs.\ \ref{F1}(a) and \ref{MSE}(a) depict the performances under the \textbf{block sparsity} pattern. The results show that algorithms explicitly designed to exploit block-sparse structure significantly outperform the standard M-SBL algorithm. This superior performance can be attributed to the ability of these specialized algorithms to effectively exploit the inherent block structure in the data during the recovery process, leading to more accurate signal reconstruction and better convergence characteristics.

The results for the \textbf{hybrid sparsity}, where multiple sparsity patterns coexist, can be found in Figs.\ \ref{F1}(b) and \ref{MSE}(b). The proposed algorithm seen to significantly outperform the M-SBL, demonstrating its adaptability to complex signal structures. In contrast, MSBL-DoL and PC-MSBL perform similarly to M-SBL, suggesting that their rigid coupling mechanisms limit their effectiveness in handling diverse sparsity patterns.

Finally, to demonstrate the flexibility and robustness of the proposed algorithm, its performance is also evaluated under \textbf{isolated sparsity} patterns. Figs.\ \ref{F1}(c) and Fig. \ref{MSE}(c)  show that the proposed algorithm achieves performance comparable to M-SBL, even in the presence of severe sparsity model mismatches. On the other hand, both PC-MSBL and MSBL-DoL exhibit significant degradation. The robustness of the proposed algorithm emerges from two key aspects of our adaptive prior $p(\gammab)$. Firstly, it learns the correlation structure directly from the data rather than imposing fixed coupling constraints. Secondly, it automatically adjusts the strength of the sparsity-promoting weights based on the observed signal statistics. This adaptive mechanism allows the algorithm to seamlessly handle both block-structured and isolated non-zero elements without requiring parameter tuning or prior knowledge of the sparsity pattern.


\section{Conclusion}
We proposed a novel SBL variant for block-sparse recovery in the MMV setup. The key innovations include an extended spatial coupling penalty that exploits block structure and a tuning-free learnable regularization weight to enhance robustness. We derived an efficient sequential closed-form ADMM-based solution for the evolved M-step. Numerical experiments demonstrate that our algorithm performs on par with state-of-the-art methods under standard sparsity assumptions and outperforms them when signals exhibit a block-sparse structure. This validates the algorithm's ability to effectively generalize to diverse signal structures. Our novel SBL variant represents an advancement in block-sparse signal recovery, with promising applications in areas such as mm-wave and near-field channel estimation.

\newpage
\bibliographystyle{IEEEtran}

\mybibliography

\end{document}